\documentclass[twocolumn,prl,aps,showpacs]{revtex4}
\usepackage{graphics}

\usepackage{dcolumn}

\usepackage{bm}

\usepackage{epsfig}
\usepackage{pictex}

\begin{document}

\title{Role of surface states in STM spectroscopy of 
(111) metal surfaces with Kondo adsorbates}
\author{J. Merino}
\affiliation{Departamento de F\'isica Te\'orica de la Materia Condensada, 
Universidad Aut\'onoma de Madrid, Madrid 28049, Spain}
\author{O. Gunnarsson}
\affiliation{Max-Planck-Institut f\"ur Festk\"orperforschung
D-70506 Stuttgart, Germany}
\date{\today}
\begin{abstract}
A nearly-free-electron (NFE) model to describe  
STM spectroscopy of (111) metal surfaces with Kondo 
impurities is presented. Surface states are found to play an important
role giving a larger contribution
to the conductance in the case of Cu(111) and Au(111) than Ag(111) surfaces. 
This difference arises from the farther extension 
of the Ag(111) surface state into the substrate.  
The different line shapes observed when Co is adsorbed on 
different substrates can be 
explained from the position of the surface band onset relative to the Fermi 
energy. The lateral dependence of the line shape amplitude is found 
to be bulk-like for $R_{||} \lesssim 4$ \AA~and surface-like at 
larger distances, in agreement with experimental data.
\end{abstract}

\maketitle

When a magnetic impurity is inside a metallic host, 
the Kondo effect can occur\cite{Kondo}. Signatures of the Kondo
effect also appear in STM measurements of   
noble metal surfaces with adsorbed Kondo impurities 
through the appearance of characteristic zero bias line 
shapes \cite{Madhavan1,Li} of the Fano-type \cite{Fano}.            
These line shapes are found to depend strongly on the
metal surface on which the magnetic atom is adsorbed.
For instance, the line shape associated with Co on Au(111) is more 
asymmetric than on Cu(111) whereas for 
Co on Ag(111) a rather symmetric line shape is observed. As both 
bulk and surface states are present in (111) surfaces 
it is yet an open question which of these play
the most important role in the Kondo effect and what 
determines the line shapes observed for the different substrates. 
  
In spite of the large amount of experimental work dedicated
to the characterization of line shapes associated with 
different noble metal surfaces a complete theoretical model for 
the tip-surface-adsorbate interaction is yet lacking.    
The interaction of bulk states in the metal with the tip and adsorbate 
has been done through a Jellium model for the surface with step
\cite{Lang,Plihal}, image \cite{Neto} and 
Jones, Jennings and Jepsen (JJJ) \cite{Jones} potentials. 
Tight-binding approaches have also been introduced\cite{Schiller}.  
However, the contribution of surface states to the conductance has
only been partially discussed \cite{Plihal,Neto}.

Previously, we studied the tip-substrate-adsorbate interaction 
considering bulk states only \cite{Merino}.  In this Letter, we introduce a NFE description of the 
substrate which is the simplest way to describe bulk and surface states 
on equal footing.  Surface states are found to give a 
substantial contribution to the STM conductance even when the tip
is right above the Kondo impurity.  This  
contribution varies from surface to surface and is larger 
for Cu(111) and Au(111) than for Ag(111). This can be understood 
from the fact that the surface state in Ag(111) extends far into the
substrate.  Finally we find that the position of the surface band onset 
relative to the Fermi energy, $\epsilon_F$, determines
the different line shapes observed in the different substrates.

A NFE description of the surface is introduced. The ionic potential 
in the (111) direction is taken into account 
as a perturbation to the Jellium potential inside the crystal. This
potential opens up a gap and surface states split from the 
bulk band edges. Since the weight of the surface states is 
removed from the bulk states one might expect that they 
are not important for integrated properties.  However, they can 
be relevant to Fermi energy properties such as the STM conductance.
Apart from including surface states we make similar assumptions as in 
Ref. \cite{Merino}: 
(i) we neglect the direct coupling of the tip with the substrate d bands
and with the 3d orbital of the adsorbate due to the localized nature of
the d orbitals \cite{Wingreen} and the large tip-surface separation,
(ii) the adsorbate is modelled by a single $d_{3z^2-r^2}$-orbital 
and the tip by a single $s$-orbital \cite{Tersoff,Gadzuk}, and
(iii) the momentum dependence (including orthogonalization effects) 
of the hybridization matrix elements are explicitly taken into 
account \cite{Grimley,Merino}.

When Co is deposited on a noble metal surface, it captures
electronic charge so that there is effectively only one hole
left \cite{Ujsaghy}. The unpaired spin residing in Co is
responsible for the Kondo effect experimentally observed.
Hence, we introduce a generalized Anderson model
to describe the substrate, adsorbate and the tip. 
Neglecting the orbital degeneracy of the 3d orbital the model 
reads
\begin{eqnarray}
H&=&\sum_{{\bf k},\sigma } \epsilon_{\bf \tilde{k}
} c^{\dagger}_{{\bf \tilde{k}}\sigma} c_{{\bf \tilde{k}}\sigma} +
\epsilon_d \sum_{\sigma} d^{\dagger}_{\sigma} d_{\sigma}
\nonumber  \\
&+& \sum_{ {\bf k},\sigma} V_{{\bf \tilde{k}}}
(d^{\dagger}_{\sigma} c_{{\bf \tilde{k}}\sigma} + H. c.)
+U d^{\dagger}_{\uparrow}
d_{\uparrow} d^{\dagger}_{\downarrow} d_{\downarrow}
\nonumber \\
&+& \sum_{{\bf k},\sigma} M_{ {\bf \tilde{k}}}
( c^{\dagger}_{{\bf \tilde{k}}\sigma } t_{\sigma} + H. c.)
+ H_{t}.
\label{hamilt}
\end{eqnarray}

Here, $\epsilon_d$ is the energy level of the adsorbate $d$ orbital, 
$c^{\dagger}_{{\bf \tilde{k}} \sigma}$ creates an electron with spin 
$\sigma$ and momentum ${\bf \tilde{k}}$. $d^{\dagger}_{\sigma}$ and 
$t^{\dagger}_{\sigma}$ create an electron in the d$_{3z^2-r^2}$ orbital 
of the adsorbate and the s-orbital of the tip, respectively.
$\epsilon_{\bf \tilde{k}}$ and  $V_{{\bf \tilde{k}}}$
are the metallic energies and the hybridization matrix elements
between the substrate and the adsorbate, respectively. $U$ is the
Coulomb repulsion of two electrons in the adsorbate.
The last two terms in the model are the tip-substrate
interaction which is governed by the matrix elements, $M_{{\bf \tilde{k}}}$, 
and $H_{t}$ that describes the tip which is supposed to
have an unstructured density of states.

The modification of the STM conductance through the surface 
due to the presence of the 3d impurity is given by \cite{Merino}
\begin{equation}
\delta G(\omega)= G_0 \rho_t
{\rm Im} \{ (A(\omega) + i B(\omega)) G_{dd}(\omega)
(A^*(\omega)+i B^*(\omega)) \},
\label{gammaAB}
\end{equation}
with $G_0={4 e^2 \over \hbar}$ and $\rho_{t}$ the tip
density of states.

In the above equation, $B(\omega)$, reads
\begin{equation}
 B(\omega)=\pi \sum_{\bf k} {M_{ {\bf \tilde{k}} } V_{ {\bf
\tilde{k}} } } \delta( \omega-\epsilon_{\bf k} ) +
\pi \sum_{\bf k_{||}} {M_{\bf \tilde{k}_{||}} } V_{ {\bf
\tilde{k}_{||} } } \delta( \omega-\epsilon_{\bf k_{||}} ),
\label{bw}
\end{equation}
where a sum over the two bulk bands appearing due to the 
ionic potential in the (111)-direction is understood in the
${\bf k}$ sum.

The first term in Eq. (\ref{bw}) describes the interaction of
the adsorbate and tip with bulk states and the second describes
the interaction with the surface band.  $A(\omega)$ is the Kramers-Kronig
transformation of $B(\omega)$.
Matrix elements, $M_{ {\bf \tilde{k}} }$ and $V_{ {\bf \tilde{k}} }$,
are evaluated with the orthogonalized wavefunctions,
$|{\bf \tilde{k}}>$. Details are given in Ref. \cite{Merino}.  
For the systems of interest here, $A(\omega)$ and $B(\omega)$ are real.   
The crucial quantity is $B(\omega)$ as it
embodies the complete information concerning the
tip-substrate-adsorbate system. $B(\omega)$ depends on the tip
position, ${\bf R}=({\bf R_{||}}, Z_t)$, the 
adsorbate position, $(0,Z_d)$, where $Z$ is refered to the 
last plane of ions.  The metal potential, $V_M$, comes in 
the matrix elements, $V_{\bf \tilde{k}}$ and $M_{\bf
\tilde{k}}$. We define throughout the paper $R_{||}=|{\bf R_{||}}|$.
The Green's function, $G_{dd}(\omega)$, describes the
electronic properties of the $3d$ adsorbate immersed in the
metallic continuum including many-body effects such as the
Kondo effect. The Kondo peak is simulated 
through a Lorentzian of width $T_K$ positioned at $\epsilon_K$
and is contained in $G_{dd}(\omega)$ with $T_K$ and $\epsilon_K$ 
taken from experimental data. These parameters vary between
$\epsilon_K \sim 3-6$ meV and $T_K \sim 50-90$ K for the
different (111) surfaces \cite{Madhavan1,Schneider,Knorr}.

Metal wavefunctions are obtained by solving Schr\"odingers 
equation in the presence of the surface potential, which
is given (in Rydberg energy units) by:
\begin{equation}
V_M(Z)=\left\{\begin{array}{lll} {-1 \over 2 (Z-Z_{im})}
(1-e^{-\lambda (Z-Z_{im})})&,\;   Z > Z_{im}
\\
\\
{-V_0 \over A e^{\beta (Z-Z_{im}) }+1 }&,\; Z_{m} <  Z < Z_{im}
\\
\\
{-V_0 + 2 V_G \cos(G Z) }&,\;Z < Z_{m} 
\end{array}\right.
\label{metpot}
\end{equation}
where $Z$ is the perpendicular direction to the surface (the (111)
direction).  $V_0$ is the depth of the bulk potential and $\lambda$
controls the sharpness of the surface barrier potential. The
parameters $A=2 V_0/\lambda -1$ and $\beta=V_0/A$ are obtained by
imposing the condition of continuous differentiability of the JJJ potential at
$Z=Z_{im}$ (see Ref. [\onlinecite{Jones}] for more details).
$Z_{m}$ is obtained by matching the periodic part of the potential
in the perpendicular direction to the surface with the JJJ potential.
All distances are referred to the last plane of ions  
and energies are referred to the vacuum level unless otherwise stated. 
$G$ is the reciprocal vector along the perpendicular direction to the
surface (the (111) direction) and $V_G$ is the first Fourier component 
of the crystal potential associated with $G$.                        
For the (111) surfaces $G=2 \pi \sqrt{3}/a$, where
$a$ is the lattice parameter.

The periodic part of the potential inside the crystal couples
the $|{\bf k}>$ and $|{\bf k+G}>$ states, leading to a gap of 
size: $E_G=2|V_G|$.  For perpendicular momenta: $0 \leq k_z \leq G/2$
and energies outside the gap, bulk wavefunctions are obtained by 
matching the wavefunctions outside and inside the crystal.    
Wavefunctions are free electron-like far from the band edges but differ
from this description close to them.

\begin{table}
\caption{Surface state data and parameters used in a NFE
model of (111) noble metal surfaces. The depth of the crystal
potential, $V_0$, the first Fourier component of the ion
potential, $V_G$, and the metal work function, $W$, are taken from
Ref. [\onlinecite{Smith}] and are given in eV's.
$V_0$ is referred to the vacuum level while the surface state energy
is referred to the Fermi energy, $\epsilon_F$.
The lattice parameter, $a$, and the decay of the surface state
into the bulk, $\lambda_{ss}$, are given in \AA~while
the steepness of the potential denoted as $\lambda$,
is given in \AA$^{-1}$.
 }
\label{table1}
\begin{tabular}{lllllllllll}
Surface & a & $V_0$ & $V_G$ & W & $\lambda$ &
$Z_{im}$ & $\epsilon_{ss}-\epsilon_F$ & $m_{ss}/m_e$ & $\lambda_{ss}$ \\
\hline
Cu(111) & 3.64  & 14.7 & 2.55 & 4.95 & 2.21 & 1.29 & -0.32 & 0.38 & 10.4 \\
Ag(111) & 4.08  & 15.1 &  2.1  & 4.74 & 2.23 & 1.28 & -0.05 & 0.39 & 28.6 \\
Au(111) & 4.07  &  17  &  2.3 & 5.30 & 2.36 & 1.25 & -0.49 & 0.26 & 13.5  \\
\hline
\end{tabular}
\end{table}

Surface states are possible inside the 
gap with momentum $k_z=G/2+i/\lambda_{ss}$. These states decay inside 
the bulk with the decay length, $\lambda_{ss}$, retaining the periodicity 
parallel to the surface. By matching the complex
solution inside the bulk with the solution outside we solve for 
the energy at which the surface state exists \cite{Smithss}.     

Wavefunctions associated with these surface states read
\begin{equation}
\Psi(\epsilon,{\bf r})= \left\{\begin{array}{lll} 
{D \over \sqrt{\sigma}} e^{i \bf{k_{||} r_{||}}} 
e^{(z-Z_{m})/\lambda_{ss}} \cos({G \over 2} z +\theta(\epsilon))  &,\;   z < Z_{m}
\\
\\
e^{i \bf{k_{||} r_{||}}} \psi(\epsilon,z) &,\; z>Z_{m} 
\\
\\
\end{array}\right.
\label{wavess}
\end{equation}
where, $\psi(\epsilon,z)$, is the solution to the Schr\"odinger
equation in the perpendicular direction for a given energy,
$\epsilon$.  $\sigma$ is the area of 
the square enclosing the surface wavefunction and $\theta(\epsilon)$ is the
phase factor coming from the solution of Schr\"odingers
equation inside the bulk. Surface states with larger $\lambda_{ss}$ 
penetrate more into the solid and correspond
to surface state energies closer to the bulk band edge. The normalization
constant, $D \sim \sqrt{1/\lambda_{ss}}$ for $\lambda_{ss} \gg 1 $, 
going to zero as the surface state energy approaches the bulk continuum.
This constant is important to determine the relative contribution of 
surface states compared to bulk states in the conductance.   

\begin{figure}
\begin{center}
\epsfig{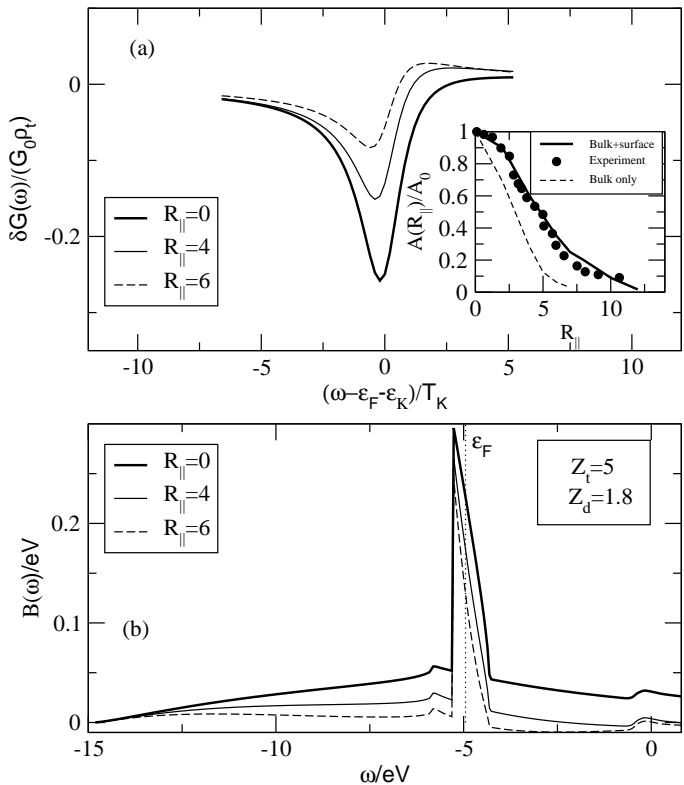}
\end{center}
\caption{Conductance line shapes obtained from the NFE model
for Co on Cu(111). In (a) we show conductance line shapes for
different lateral positions of the tip associated with
the function $B(\omega)$ displayed in (b).  The inset shows the lateral
variation of the normalized conductance amplitude, $A(R_{||})/A_0$ 
compared to experimental data. Distances are given in \AA.
}
\label{fig1}
\end{figure}

The surface state energy is determined   
by keeping all parameters describing the surface potential 
fixed except for $Z_{im}$ which is varied to recover 
the surface state position in agreement with photoemission 
data [\onlinecite{Smith}]. 
The values of $Z_{im}$ obtained are shown in Table \ref{table1}.
For instance, for Cu(111) a 
surface state at 0.32 eV appears below the Fermi energy and the first image state at 
0.83 eV below the vacuum level for $Z_{im}=1.29$ \AA~ 
in good agreement with experimental values.  Bulk-band effective masses,
$m^*/m_e=0.74$, 0.55, and 0.52 are taken
in the (111) direction and 0.25, 0.36 and 0.18 in the parallel 
direction for Cu, Ag and Au, respectively.
In our calculations of $B(\omega)$ we neglect the image state band
as it is far from the Fermi energy and disperses over a rather small
energy range, leading to a small contribution to $A(\epsilon_F)$. 
  
Parameters describing surface bands of Cu(111), Au(111) and
Ag(111) surfaces are summarized in Table \ref{table1}, where 
relevant parameters of the surface potential
together with values of $\lambda_{ss}$ and surface band 
effective masses, $m_{ss}$,  are displayed. These masses are 
in good agreement with 
available experimental data from STM of clean surfaces 
\cite{Fiete,Schneider,Chen} and photoemission data \cite{Smith}.  

\begin{figure}
\begin{center}
\epsfig{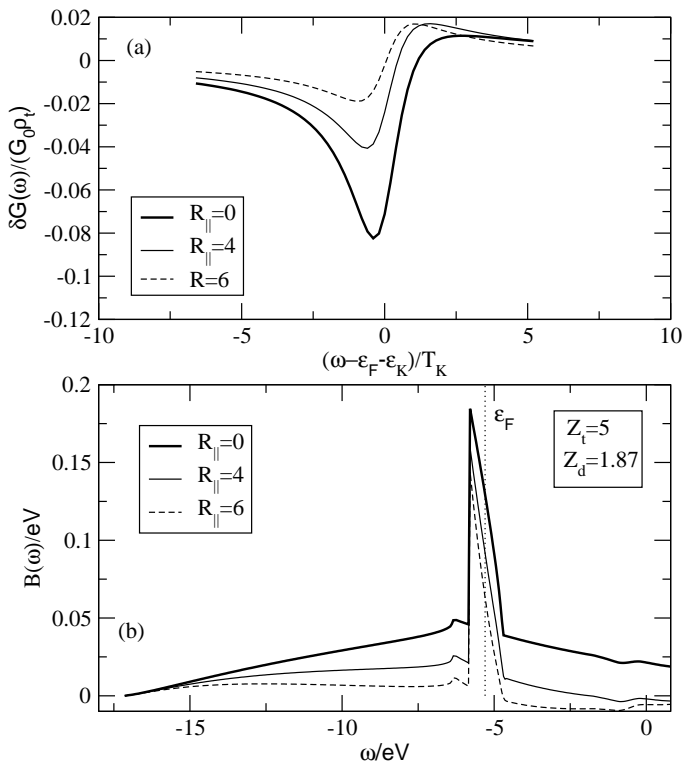}
\end{center}
\caption{Conductance line shapes calculated from the
NFE model for Co on Au(111). Line shapes become more
asymmetric with more positive Fano parameter as 
$\epsilon_{F}-\epsilon_{ss}$ increases (compare
the plotted line shape for Au(111) with the ones
shown in Figs. \ref{fig1} and \ref{fig3}).
}
\label{fig2}
\end{figure}

\begin{figure}
\begin{center}
\epsfig{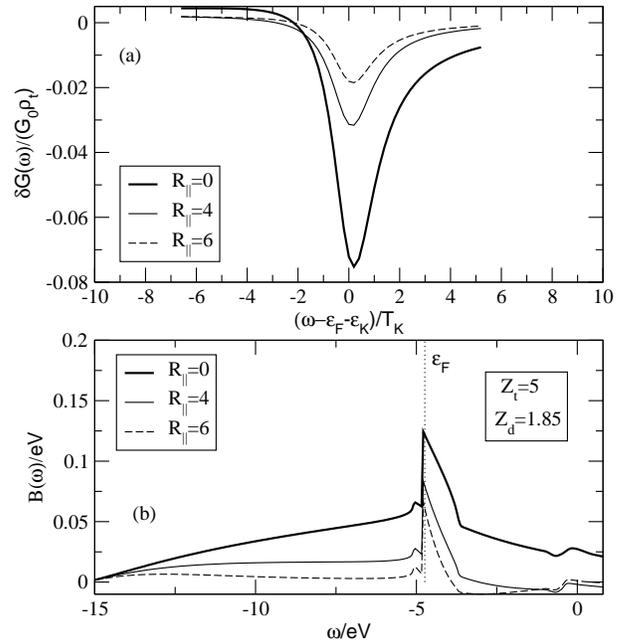}
\end{center}
\caption{Conductance line shapes calculated from the
NFE model for Co on Ag(111). The surface band gives the
smallest contribution of the three noble metal surfaces
to $B(\omega)$ and the conductance.  }
\label{fig3}
\end{figure}

The conductance for  
the three noble metal surfaces with Co adsorbed on them is calculated. 
Results for $B(\omega)$ and conductance line shapes
are displayed in Fig. \ref{fig1} to Fig. \ref{fig3} 
for Cu, Au and Ag, respectively. Most of the contribution
to $B(\omega)$ close to the Fermi energy comes from the
surface band and is larger for Cu(111) and
Au(111) than for Ag(111) as expected as in this case the
surface state penetrates more into the crystal.  At the same time $B(\omega)$ 
has a rapid drop with increasing $\omega$. This is because as the 
surface band disperses it gradually approaches the bulk continuum so that
at a certain energy the surface band hits the bulk band and its  
amplitude drops to zero. 

Depending on the position of the surface band onset
relative to the Fermi energy, the conductance line shape becomes more or
less symmetric. If there is more weight below the Fermi energy,
the conductance line shape is asymmetric with positive Fano parameter,
 $q>0$ (see Fig. 1 of Ref.\cite{Merino}). From Table. \ref{table1} we see that $|\epsilon_{ss}-\epsilon_F|$
is smallest for Ag(111) and largest for Au(111). Hence,
the most asymmetric line shape corresponds to Au(111) and the
most symmetric one to Ag(111). The different line shapes
calculated for the different substrates are in agreement
with experimental observations. 

The conductance amplitude, $A(R_{||})$, 
is more rapidly suppressed at short distances $R_{||} \lesssim 4$ \AA~than for $R_{||} > 4$ \AA.  
This is because both bulk and 
surface states contribute to $B(\epsilon_F)$.  As the tip 
is displaced laterally the bulk contribution to $B(\omega)$ close
to $\epsilon_F$ drops quickly until it eventually vanishes.  Farther apart from the impurity,  
the surface band dominates and the lateral dependence becomes surface-like.  
Had we considered bulk states only in our model,  
$A(R_{||}=6)/A_0 \sim 3 \%$,    
which is an order of magnitude smaller than the 
one found experimentally\cite{Knorr}. This is because $A(R_{||}) \sim
1/R_{||}^2$ for bulk states. In contrast, the  
amplitude dependence on $R_{||}$  
is in good agreement with our calculations when both surface and bulk states are 
included (see inset of Fig. \ref{fig1}).

In conclusion, surface bands give an important
contribution to the conductance in STM measurements
of (111) noble metal surfaces with Kondo adsorbates.
The position of the surface band onset 
relative to the Fermi energy determines the line shapes 
in the different (111) substrates. 
The lateral variation of the amplitude is found to be  
bulk-like close to the adsorbate becoming surface-like 
for $R_{||} > 4$ \AA~in agreement with experimental observations.

\acknowledgments
We acknowledge helpful discussions with Y. Dappe, 
P. Jelinek, M. A. Schneider and P. Wahl. M. A. S. and P. W. for 
lending their experimental data to us.  J. M. acknowledges 
financial support from 
the Ram\'on y Cajal program from Ministerio de Ciencia y 
Tecnolog\'ia in Spain.

\end{document}